# Chiral tunneling in trilayer graphene

S. Bala Kumar and Jing Guo[a)]

Department of Electrical and Computer Engineering, University of Florida, Gainesville, FL, 32611

[*]Corresponding author. E-mail: a) guoj@ufl.edu

Abstract

We study the effect of chiral-tunneling in Bernal and Rombhohedral stacked trilayer-graphene (3LG). Based on the chirality of the electronic bands, at the K-point, (Rombhohedral) Bernal-3LG exhibits 100% (50%) transparency across a heterojunction. Utilizing this property, we further investigate the effect of electron collimation in 3LG. Due to the difference in the Berry's phase, we show that, Rombhohedral-3LG is a better electron collimator, compared to monolayer and Bernal-bilayer graphene. Since, Bernal-3LG can be decomposed into two separate channels consisting of a monolayer and a modified Bernal-bilayer graphene; the Bernal-3LG is weaker electron collimator, compared to Rombhohedral-3LG.

Graphene is a two dimensional material with many unusual properties[1, 2], which could be useful in designing future electronic devices[3-5]. The chiral nature of electron's wave function in graphene, gives rise to unique electronic behavior, known as chiral tunneling[6-8], where a normally incident electron is transparent across a potential barrier. Chiral tunneling effect has also been experimentally observed[9, 10]. On the other hand, recently many studies are being conducted on electronic transport in trilayer graphene (3LG)[11-17]. 3LG has two natural stable allotropes: 1) Bernal (ABA) stacking and 2) Rombhohedral (ABC) stacking.[18] Even though, simply by shifting the top layer of an ABA-3LG, the ABC-3LG is obtained, both ABA-3LG and ABC-3LG have remarkably different physical properties.[19-21] For example, near the Dirac point, the electrons in ABC-3LG behaves as massive fermions, while in ABA-3LG electrons behave as both massive and massless Dirac fermions[21-25]. Due to these differences, we can expect ABA-3LG and ABC-3LG to exhibit interesting variations in terms of chiral tunneling.

In this letter, using π-orbital tight binding model (π-TB), we investigate the effect of chiral tunneling in both the ABA-3LG and ABC-3LG. At K-point, we find that the ABC-3LG is transparent across a PN-junction. We derive a general dirac Hamiltonian for an n-layer graphene and show that, all odd (even) layered Rombhohedral graphenes are transparent (completely reflective) at the K-point. On the other hand, ABA-graphene can be decomposed into two separate channels consisting of a monolayer-graphene (1LG) and a modified AB-bilayer-graphene (2LG). The 1LG channel is transparent, while the AB-2LG channel is reflective, and thus at the K-point only 50% of the incoming electrons are transmitted across the PN-junction. We further investigate the effect of electron injection-angle on the chiral tunneling. Our results show that ABC-3LG is the best electron collimator, compared to ABA-3LG, as well as 1LG and 2LG.

We used nearest neighbor (NN)[20] π-TB model to investigate the electronic transport across the graphene layers. The π-TB Hamiltonian for a 1LG[26] at K-point is

$$h(K_x,K_y) = \begin{bmatrix} 0 & -t_0 - 2t_0\cos(K_y - 2\pi/3)e^{iK_x} \\ -t_0 - 2t_0\cos(K_y - 2\pi/3)e^{-iK_x} & 0 \end{bmatrix} \quad (1)$$

where $K_y = \sqrt{3}k_y a_0/2$, $K_x = 3k_x a_0/2$, intralayer NN hoping parameter, $t_0$=0.27eV, and the NN intralayer atomic distance $a_0$=0.142nm. Equation 1 can be expressed as follows

$$h(K_x,K_y) \equiv \alpha + \beta(K_y)e^{+iK_x} + \beta(K_y)^\dagger e^{-iK_x}, \quad (2)$$

where $\alpha = \begin{bmatrix} 0 & -t_0 \\ -t_0 & 0 \end{bmatrix}$, and $\beta(K_y) = \begin{bmatrix} 0 & -2t_0\cos(K_y - 2\pi/3) \\ 0 & 0 \end{bmatrix}$. Using Eq. 2, a 2D-graphene layer can be model as a chain of 2x2 matrices for any given $K_y$ value. As illustrated in Fig. 1(a), $\alpha$ is the onsite matrix and $\beta(K_y)$ is the hopping matrix. For PN-junction, the electric potential U(x) is added to the onsite energy. For the 3LG, the structure of the first row of the ABA-3LG and ABC-3LG is shown in Fig. 1(b-c). The interlayer NN hopping parameter, $t_p$=0.35eV. Figure 1 (d) and (e) shows the band structure at the K-point for the ABC-3LG and ABA-3LG, respectively. Using the non-equilibrium Green's function (NEGF) formalism[26], we compute the electron transmission at Fermi energy, $E_F$ for a given $K_y$, $T(E_F,K_y)$. The conductance further computed as $g = g_0 \int_{-\pi}^{+\pi} T(E_F,K_y)dK_y$, where $g_0 = e^2/h$.

The $T(E_F,K_y)$ map for ABA-3LG and ABC-3LG is plotted in Fig. 2. At low energy of $E_F < t_p$, when the electron energy is larger than the barrier height, i.e. $E_F > U_0$, ABC-3LG has a maximum transmission of T=1 due to one lowest (highest) conduction (valence) band, whereas the maximum transmission of an ABA-3LG is T=2 due to two lowest (highest) conduction (valence) band. In conventional semiconductors, when the $E_F<U_0$, the transmission T decays to zero across the PN-junction. However, due to chiral tunneling, we obtain finite transmission across the junction. For example, at $K_y=0$, the ABA (ABC) -3LG has a transparency of 50% (100%), with transmission of T=1.

First we explain the observation for the ABC-3LG. By using small energy approximation, we derive the Dirac Hamiltonian for n-layer Rombhohedral-stacked graphene

$$H(k) = \frac{(\hbar v k_F)^n}{t_p^{n-1}} \begin{bmatrix} 0 & e^{-in\theta} \\ e^{+in\theta} & 0 \end{bmatrix}, \qquad (3)$$

where $k_F = \sqrt{k_x^2 + k_y^2}$ and $\theta = \tan^{-1}(k_y/k_x)$. From Eq. (3) the eigenenergy, $E_F$ and corresponding wavefunction, $\psi$ of the lowest (highest) conduction (valence) band is

$$E_F = \pm \frac{(\hbar v k_F)^n}{t_p^{n-1}} \Rightarrow \psi = \begin{bmatrix} \pm e^{in\theta} \\ 1 \end{bmatrix}. \qquad (4)$$

Note that for n-layer graphene the Berry phase is $n\pi$ and $E_F \propto (k_F)^n$. This approximation is valid for small energy, and the quantitative deviation of this approximation gets more significant as the n increases.

Referring to Fig. 3(a), we label the wavefunction of the conduction (valence) band with positive momentum as $\psi_C(\psi_V)$. Across the PN-junction, electrons are tunneled from the conduction to the valence band. Therefore finite transmission is obtained when the wavefunctions are not orthogonal, i.e. $\gamma = \langle \psi_C | \psi_V \rangle \neq 0$. When ky=0, the θ=0 and thus $E_F = \pm \frac{(\hbar v k_x)^n}{t_p^{n-1}} \Rightarrow \psi = \begin{bmatrix} \pm 1 \\ 1 \end{bmatrix}$. A careful analysis of this equation shows that for any odd (even) layered graphene stacked in ABC configuration, the $\psi_C$ is equal (orthogonal) to $\psi_V$, resulting in a perfect transmission (reflection) across the barrier. We have clearly illustrated this in Fig. 3(b-d). Therefore ABC-3LG, being odd layered, has a perfect transmission T=1, across the barrier, at $K_y$=0.

Next we analyze the results obtained for the ABA-3LG in Fig. 2(b). The Hamiltonian of ABA-3LG in the basis vectors <1|, <2|, <3|, <4|, <5|, and <6| as labeled in the inset of Fig. 1(e) is

$$H_{ABA'} = \begin{bmatrix} 0 & \hbar v_F k_F e^{-i\theta} & 0 & t_p & 0 & 0 \\ \hbar v_F k_F e^{+i\theta} & 0 & 0 & 0 & 0 & 0 \\ 0 & 0 & 0 & \hbar v_F k_F e^{-i\theta} & 0 & 0 \\ t_p & 0 & \hbar v_F k_F e^{+i\theta} & 0 & t_p & 0 \\ 0 & 0 & 0 & t_p & 0 & \hbar v_F k_F e^{-i\theta} \\ 0 & 0 & 0 & 0 & \hbar v_F k_F e^{+i\theta} & 0 \end{bmatrix}$$

This Hamiltonian can be rewritten in the basis of $\frac{\langle 1| + \langle 5|}{\sqrt{2}}, \frac{\langle 2| + \langle 6|}{\sqrt{2}}, \langle 3|, \langle 4|, \frac{\langle 1| - \langle 5|}{\sqrt{2}}, \frac{\langle 2| - \langle 6|}{\sqrt{2}}$ as follows

$$H_{ABA} = \begin{bmatrix} 0 & \hbar v_F k_F e^{-i\theta} & 0 & \sqrt{2} t_p & 0 & 0 \\ \hbar v_F k_F e^{+i\theta} & 0 & 0 & 0 & 0 & 0 \\ 0 & 0 & 0 & \hbar v_F k_F e^{-i\theta} & 0 & 0 \\ \sqrt{2} t_p & 0 & \hbar v_F k_F e^{+i\theta} & 0 & 0 & 0 \\ 0 & 0 & 0 & 0 & 0 & \hbar v_F k_F e^{-i\theta} \\ 0 & 0 & 0 & 0 & \hbar v_F k_F e^{+i\theta} & 0 \end{bmatrix} \equiv \begin{bmatrix} H_{AB'} & 0 \\ 0 & H_A \end{bmatrix}$$

Note that $H_{AB'}$ is the Hamiltonian of a AB-2LG with modified hopping parameter of $\sqrt{2}t_p$, while $H_A$ is the Hamiltonian of a 1LG. Therefore at low energy, ABA-3LG can be modeled as two independent channels consisting of: (1) a modified-AB-2LG (AB'-2LG) and (2) a 1LG. The AB'-2LG (1LG) channel is responsible for the ABA-3LG's parabolic (linear) band. Referring to Fig. 3(b-c), across the PN-junction only one of the two bands is transparent, i.e. the linear (parabolic) band corresponding to 1LG (AB'-2LG) is transparent (reflective). Therefore there is a 50% transparency across the barrier.

We further investigate the dependence of the transmission across the PN-junction, on the electron injection-angle, $\theta$. In Fig. 4, we show results for 1LG, 2LG and 3LGs. For 1LG, AB-2LG, and ABC-3LG the $k_F$ for a given $E_F$ is estimated from Eq. (3) as $k_F^1 = \frac{E_F}{\hbar v}$, $k_F^{AB} = \frac{\sqrt{t_p E_F}}{\hbar v}$, and $k_F^{ABC} = \frac{\sqrt[3]{t_p^2 E_F}}{\hbar v}$, respectively. Since, ABA-3LG has two $k_F$ values corresponding to the two bands, and we use the larger $k_F$ value, i.e $k_F$ of the parabolic band with modified hopping parameter, i.e. $k_F^{ABA} = \frac{\left(2t_p^2 E_F\right)^{1/3}}{\hbar v}$. From this $k_F$ and $\theta$, we compute the ky and then using the $\pi$-TB model we plot the transmission $T(\theta, E_F)$ as a function of $\theta$ for different $E_F$ values [Fig. 4].

The transmission when electron energy $E_F = U_0/2$, is shown by the thick curves in Fig. 4. At $E_F = U_0/2$, the $k_F$ of the injected electrons is equal to the $k_F$ of the transmitted electrons. At this energy, for the 1LG there is a finite transmission at all angles $\theta$, whereas for AB-2LG there is a forbidden angle, at $\theta = 0^0$, where electrons are complete reflected. On the other hand, ABC-3LG has two forbidden angles, at $\theta = \pm 30^0$, and most transmission occurs at a within angles of $|\theta| < 30^0$.

Since, the transmission of the ABA-3LG is an addition of 1LG and AB'-2LG, and we obtain finite transmission at all the angles θ, as in 1LG. Therefore, as showed by the results in Fig. 4, ABC-3LG is the best electron collimator, in comparison to 1LG, AB-2LG, and ABA-3LG. For n-layer graphene, the forbidden $\theta = -\pi/2 + j\pi/n$ for $j = 1, 2, \ldots, n-1$. Note that for normal injection, i.e. θ=0, transmission is forbidden (allow) for even (odd) layered graphenes. We also plotted the electron transmission across a rectangular barrier, as shown by the thin lines in Fig. 4. The same conclusion as in transmission across PN-junction remains i.e. ABC-3LG is the best electron collimator. We also verified the above results in a more realistic barrier, where the potential varies smoothly, and find that the find that the qualitative conclusion on electron collimation remains the same.

In conclusion, we studied the chiral tunneling effect in 3LG. At $k_y$=0, ABC-graphene is 100% transparent across a PN-junction. ABA-3LG has two bands at low energy: linear band with a similar behavior as the 1LG and a parabolic band with similar behavior as AB-2LG. Across a PN-junction, at ky=0, the linear band is transparent, while the parabolic band is reflective. We further studied the effect of injection-angle, θ on chiral-tunneling. Our results indicate that ABC-3LG has the highest degree of electron collimation.

**Figures and Caption**

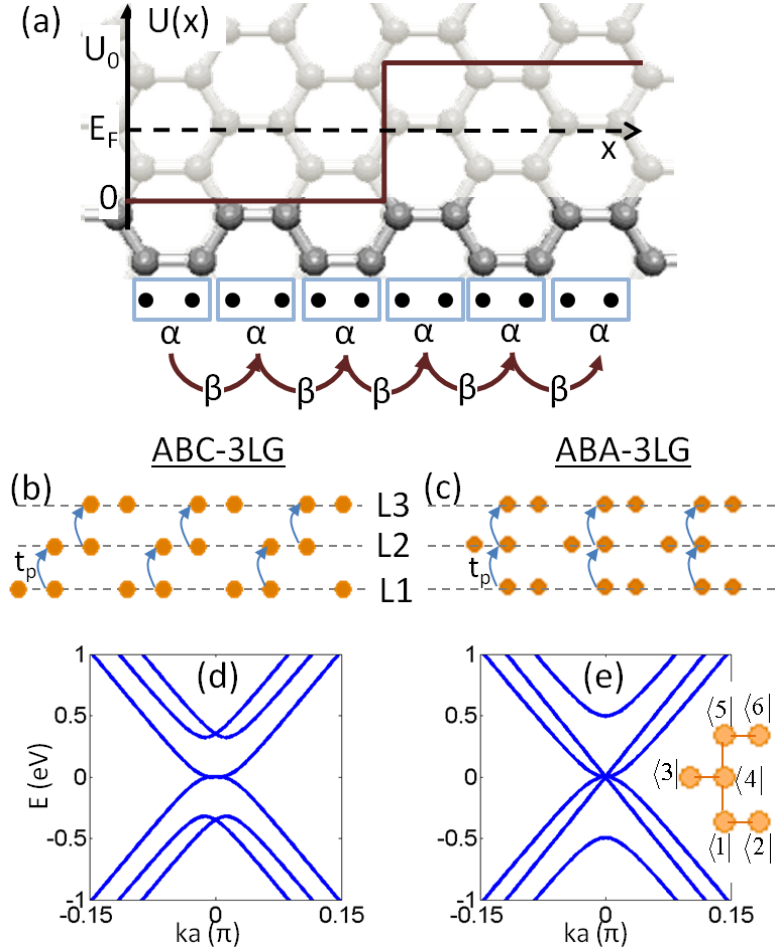

Figure 1 (a) PN-junction barrier across a layer of graphene. The graphene layer is modeled as a chain of 2x2 matrix with α (β) as the onsite (hopping) matix. The electric potential U(x) is added to the onsite energy. The side view of the first row of a (b) ABC-3LG and (c) ABA-3LG. The curved-arrows indicate the NN interlayer coupling [$t_p$=0.35eV]. Band structure of the (d) ABC-3LG and (e) ABA-3LG at $k_y$=0. The inset in (e) shows the basis vector for an ABA-3LG unit cell.

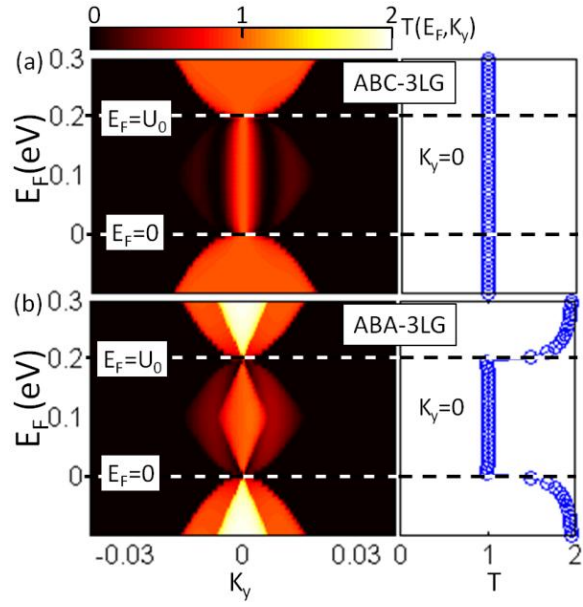

Figure 2 The $T(E_F, K_y)$ map for (a) ABC-3LG and (b) ABA-3LG. The right panels show the transmission at Ky=0. For ABA-3LG the transmission decreases 50%, from T=2 to T=1 across a PN-junction with barrier height $U_0$=0.2eV. For ABC-3LG, the transmission remains T=1 both within and outside the barrier.

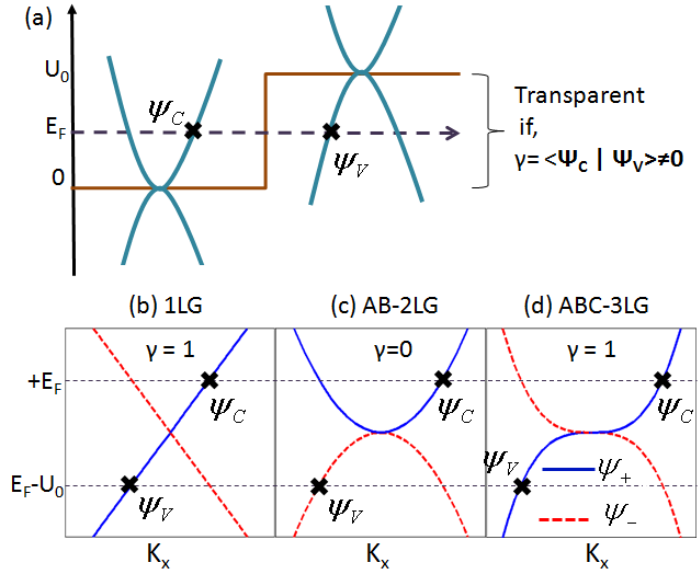

Figure 3 (a) Tunneling across the PN-junction. The PN-junction is transparent if $\gamma \neq 0$. Band structure of the lowest (highest) conduction (valence) band for (b) 1LG, (c) AB-2LG, and (d) ABC-3LG at the K-point. The $\psi_+ = [1,1]^\dagger$ and $\psi_- = [-1,1]^\dagger$ are orthogonal to each other, and thus at $E_F$, the $\gamma = 0$ ($\gamma = 1 \neq 0$) for AB-2LG (1LG and ABC-3LG).

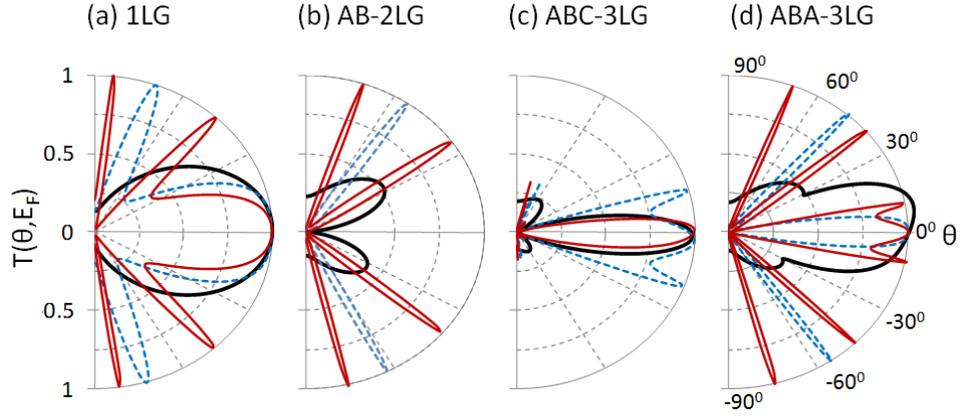

Figure 4 Dependence of transmission, $T(\theta, E_F)$ on the electron injection-angle, $\theta$, for (a) 1LG, (b) AB-2LG, (c) ABC-3LG, and (d) ABA-3LG. The thick curves are transmission across PN-junction with $U_0=0.2$eV and $E_F=U_0/2$. The thin curves are transmission across a 100nm wide rectangular (NPN) -barrier at electron energy, (a) $E_F=50$meV (b-d) $E_F=25$meV. The NPN-barrier heights are: (a) 100 and (b-d) $\approx 65$meV (solid-thin curves); and (a) 120 and (b-d) $\approx 80$meV (dotted-thin curves); and. Note that the curves in (d), are similar to that obtain by adding the transmission of 1LG and AB'-2LG using $k_F = k_F^{ABA}$. The thin lines in (a) and (b) are qualitatively similar to the results obtained by M. I. Katsnelson et al. (see Ref. 6).

# References

1  A. K. Geim and K. S. Novoselov, Nature Materials **6**, 183 (2007).
2  A. H. Castro Neto, F. Guinea, N. M. R. Peres, K. S. Novoselov, and A. K. Geim, Reviews of Modern Physics **81**, 109 (2009).
3  N. S. Norberg, G. M. Dalpian, J. R. Chelikowsky, and D. R. Gamelin, Nano Letters **6**, 2887 (2006).
4  K. S. Novoselov, D. Jiang, F. Schedin, T. J. Booth, V. V. Khotkevich, S. V. Morozov, and A. K. Geim, Proceedings of the National Academy of Sciences of the United States of America **102**, 10451 (2005).
5  C. Berger, Z. Song, T. Li, X. Li, A. Y. Ogbazghi, R. Feng, Z. Dai, A. N. Marchenkov, E. H. Conrad, P. N. First, and Walt A. de Heer, The Journal of Physical Chemistry B **108**, 19912 (2004).
6  M. I. Katsnelson, K. S. Novoselov, and A. K. Geim, Nat Phys **2**, 620 (2006).
7  C. W. J. Beenakker, Reviews of Modern Physics **80**, 1337 (2008).
8  V. V. Cheianov and V. I. Fal'ko, Physical Review B **74**, 041403 (2006).
9  N. Stander, B. Huard, and D. Goldhaber-Gordon, Physical Review Letters **102**, 026807 (2009).
10 A. F. Young and P. Kim, Nat Phys **5**, 222 (2009).
11 F. Zhang, B. Sahu, H. Min, and A. H. MacDonald, Physical Review B **82**, 035409 (2010).
12 M. Koshino, Physical Review B **81** (2010).
13 M. F. Craciun, S. Russo, M. Yamamoto, J. B. Oostinga, A. F. Morpurgo, and S. Thrucha, Nature Nanotechnology **4**, 383 (2009).
14 W. Bao, L. Jing, J. Velasco Jr, Y. Lee, G. Liu, D. Tran, B. Standley, M. Aykol, S. B. Cronin, D. Smirnov, M. Koshino, E. McCann, M. Bockrath and C. N. Lau, Nat Phys **7**, 948 (2011).
15 T. Taychatanapat, K. Watanabe, T. Taniguchi, and P. Jarillo-Herrero, Nat Phys **7**, 621 (2011).
16 S. B. Kumar and J. Guo, Applied Physics Letters **98**, 222101 (2011).
17 C. H. Lui, Z. Li, K. F. Mak, E. Cappelluti, and T. F. Heinz, Nat Phys **7**, 944 (2011).
18 H. Lipson and A. R. Stokes, Proc. R. Soc. London **A181**, 101 (1942).
19 M. Aoki and H. Amawashi, Solid State Communications **142**, 123 (2007).
20 F. Guinea, A. H. Castro Neto, and N. M. R. Peres, Physical Review B **73**, 245426 (2006).
21 S. Latil and L. Henrard, Physical Review Letters **97**, 036803 (2006).
22 C. L. Lu, C. P. Chang, Y. C. Huang, R. B. Chen, and M. L. Lin, Physical Review B **73**, 144427 (2006).
23 B. Partoens and F. M. Peeters, Physical Review B **74**, 075404 (2006).
24 F. Guinea, Physical Review B **75**, 235433 (2007).
25 M. Koshino and E. McCann, Physical Review B **80**, 165409 (2009).
26 S. Datta, *Quantum Transport: Atom to Transistor* (Cambridge University Press, Cambridge, Cambridge University Press,).